\documentclass[aps,prl,twocolumn,superscriptaddress,showpacs,amsmath,graphicx,amsfonts,amssymb,multirow,reprint,nofootinbib]{revtex4-1}
\usepackage{graphicx}
\usepackage{siunitx}
\usepackage{latexsym}
\usepackage{amsmath}
\usepackage{txfonts}
\usepackage{helvet}
\usepackage{braket}
\usepackage{amscd}
\usepackage{longtable}
\usepackage{epsfig}

\newcommand{\BS}{\boldsymbol{\sigma}}
\newcommand{\BD}{\boldsymbol{\delta}}
\newcommand{\subPNC}{\rm{\tiny{PNC}}}
 
\begin{document}

\title{Measuring molecular parity nonconservation using nuclear magnetic resonance spectroscopy}
\author{J. Eills}
\email{eills@soton.ac.uk}
\affiliation{School of Chemistry and Institute for Life Sciences, University of Southampton, Southampton, United Kingdom}
\affiliation{Johannes Gutenberg-Universit{\"a}t  Mainz, 55128 Mainz, Germany}
\author{J. W. Blanchard}
\email{blanchard@uni-mainz.de}
\affiliation{Helmholtz-Institut Mainz, 55099 Mainz, Germany}
\author{L. Bougas}
\affiliation{Johannes Gutenberg-Universit{\"a}t  Mainz, 55128 Mainz, Germany}
\author{M. G. Kozlov}
\affiliation{Petersburg Nuclear Physics Institute, Gatchina, 188300, Russia}
\affiliation{St. Petersburg Electrotechnical University “LETI”, Professor Popov
Street 5, St. Petersburg 197376, Russia}
\author{A. Pines}
\affiliation{Materials Science Division, Lawrence Berkeley National Laboratory, Berkeley, CA, 94720 USA}
\affiliation{Department of Chemistry, University of California at Berkeley, CA, 94720 USA}
\author{D.Budker}
\affiliation{Johannes Gutenberg-Universit{\"a}t  Mainz, 55128 Mainz, Germany}
\affiliation{Helmholtz-Institut Mainz, 55099 Mainz, Germany}
\affiliation{Department of Physics, University of California, Berkeley, CA 94720-7300 USA}
\affiliation{Nuclear Science Division, Lawrence Berkeley National Laboratory, Berkeley, CA 94720 USA}

\date{\today}

\begin{abstract}
The weak interaction does not conserve parity and therefore induces energy shifts in chiral enantiomers that should in principle be detectable in molecular spectra. Unfortunately, the magnitude of the expected shifts are small and in spectra of a mixture of enantiomers, the energy shifts are not resolvable. We propose a nuclear magnetic resonance (NMR) experiment in which we titrate the chirality (enantiomeric excess) of a solvent and measure the diasteriomeric splitting in the spectra of a chiral solute in order to search for an anomalous offset due to parity nonconservation (PNC). We present a proof-of-principle experiment in which we search for PNC in the \textsuperscript{13}C resonances of small molecules, and use the \textsuperscript{1}H resonances, which are insensitive to PNC, as an internal reference. We set a new constraint on molecular PNC in \textsuperscript{13}C chemical shifts at a level of $10^{-5}$\,ppm, and provide a discussion of important considerations in the search for molecular PNC using NMR spectroscopy.
\end{abstract}



\maketitle

\section{Introduction}
\begin{figure}
  \includegraphics[width=\columnwidth]{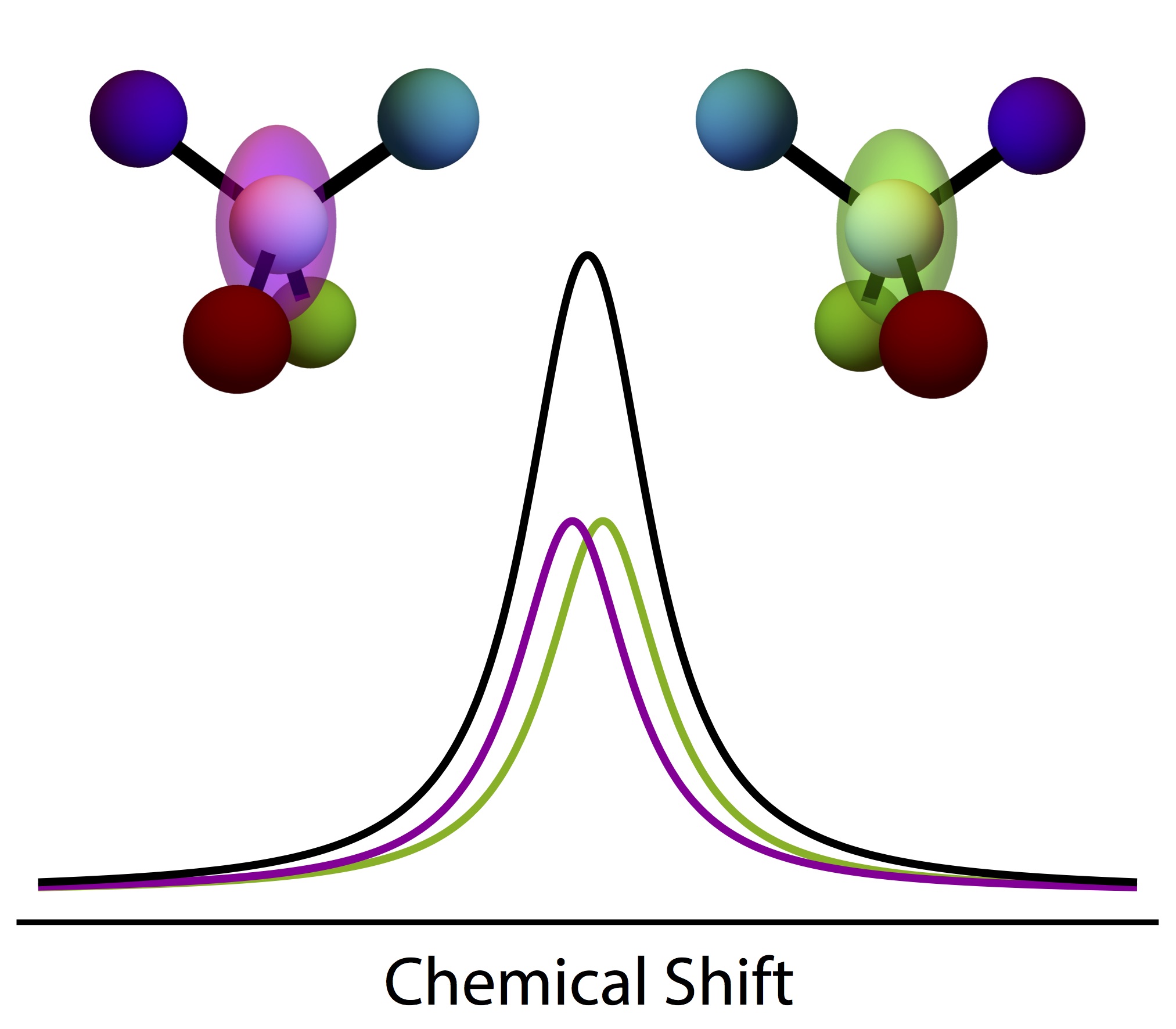}
  \caption{Illustration of the parity-violating chemical shift difference between the chiral nuclei of two enantiomers in an NMR spectrum. Coupling of the nuclear PNC terms to the electron wavefunctions breaks the symmetry of the chemical shift tensors. This diminutive chemical shift difference is too small to be observed directly. The black line represents the single peak we would expect to observe in a real spectrum.}
  \label{Fig1}
\end{figure}

\begin{figure}[h]
  \includegraphics[width=\columnwidth]{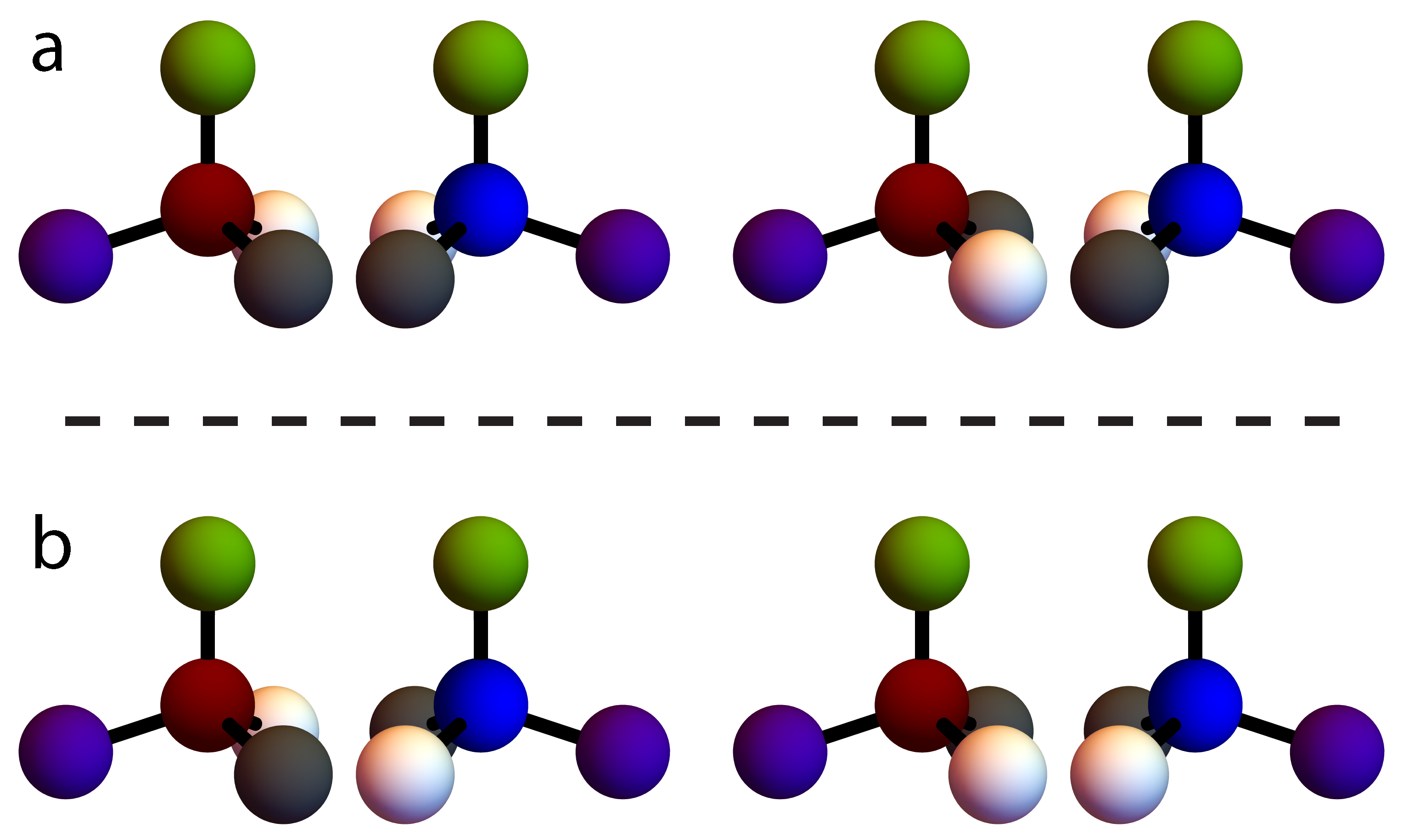}
  \caption{a) A representation of the two diastereomers that can form after a chiral solvent, the blue-center molecule, is added to a mixture of chiral enantiomers (red-center). Before the solvent is introduced, the nuclear sites of the red enantiomers give the same chemical shift in an NMR spectrum. After the chiral solvent is added, two distinct diastereomers are formed, which show different chemical shifts for the same nuclear site: a diastereomeric splitting. b) By introducing the second enantiomer of the solvent, each enantiomer of the red-center molecules exhibits a time-averaged chemical shift between the two solvent molecules, and, ignoring PNC effects, the diastereomeric splitting vanishes.}
  \label{Fig8}
\end{figure}

\begin{figure}
  \includegraphics[width=\columnwidth]{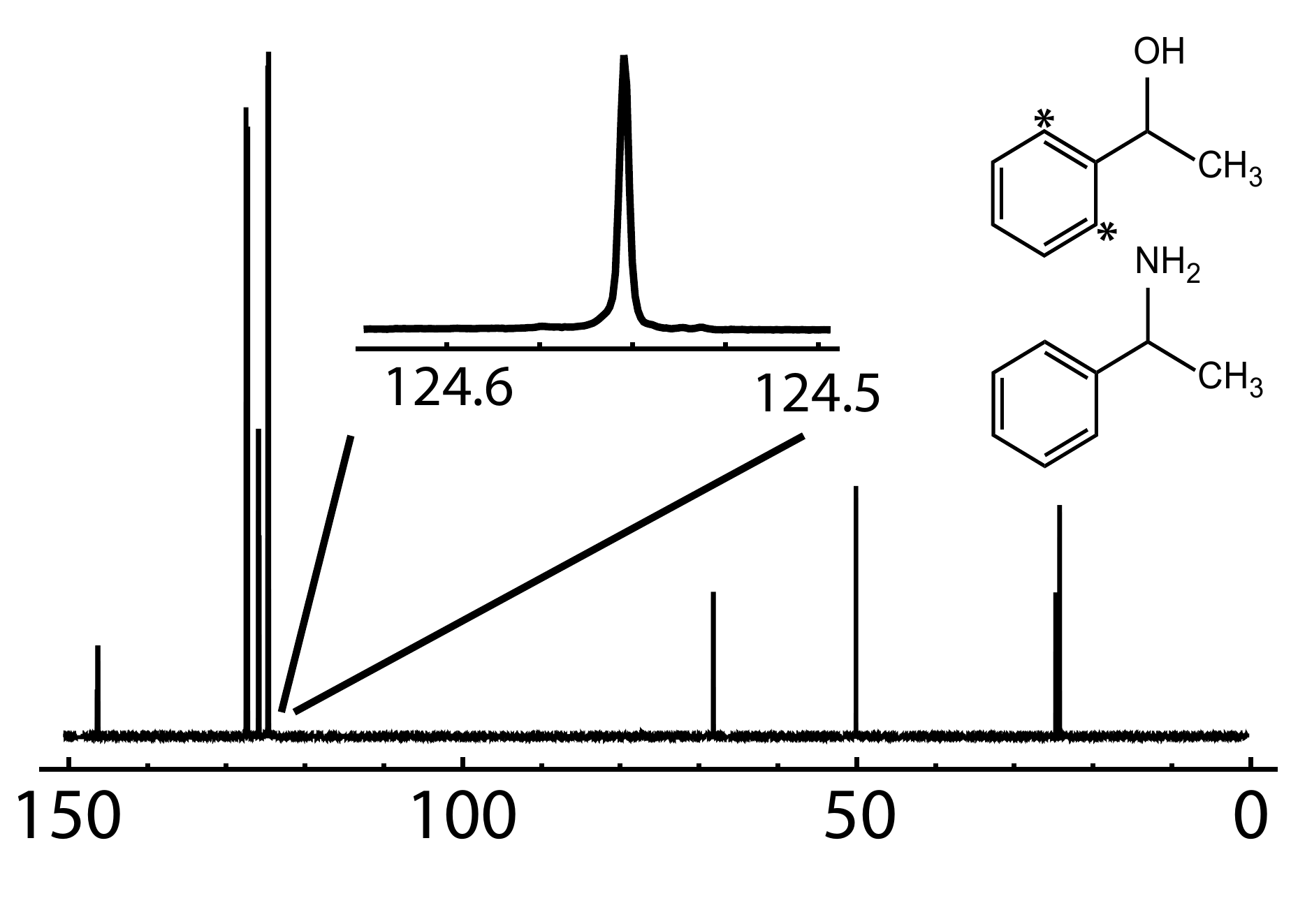}
  \includegraphics[width=\columnwidth]{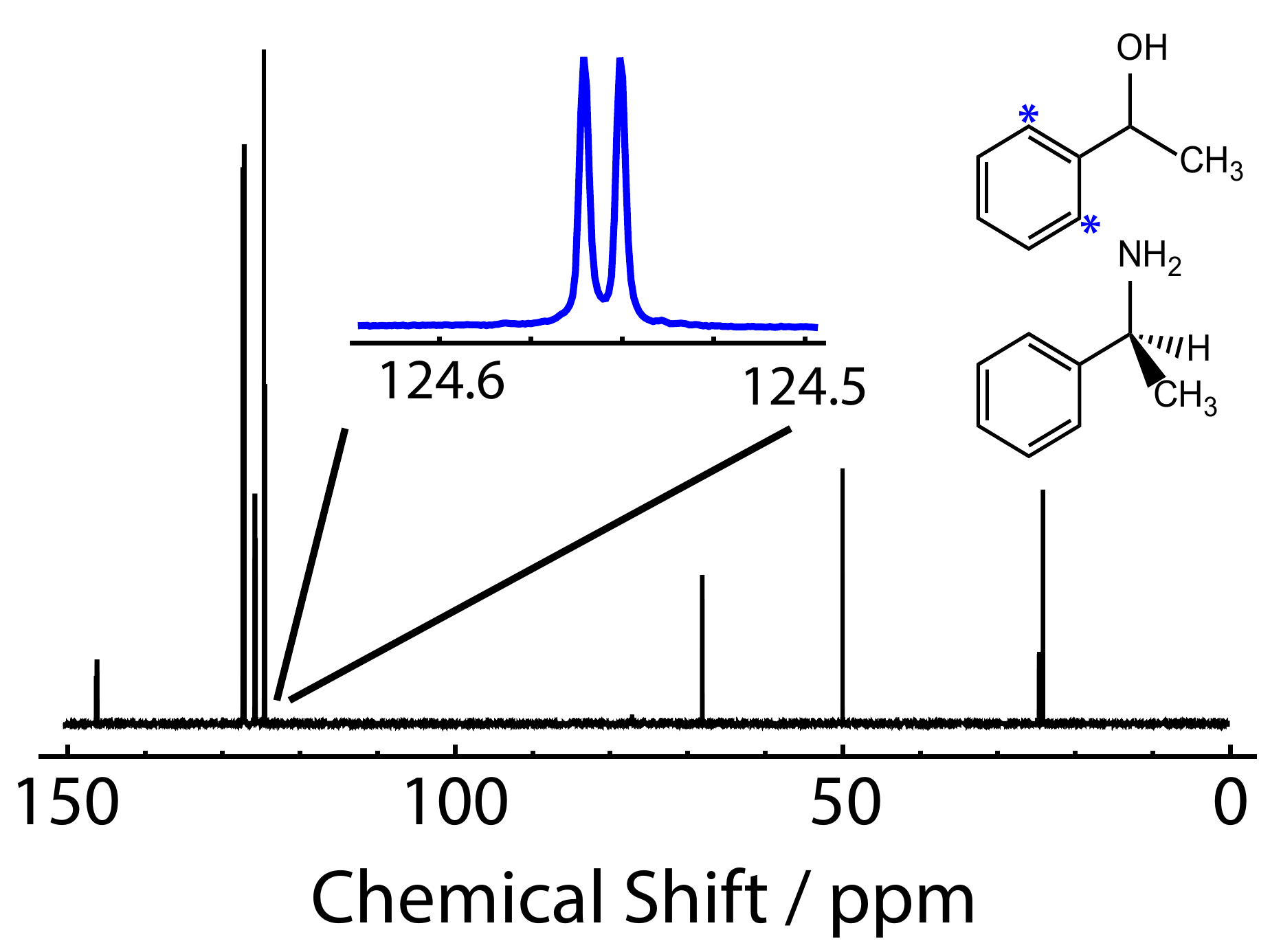}
  \caption{Proton-decoupled natural abundance \textsuperscript{13}C spectra of racemic 1-phenylethanol dissolved in CDCl\textsubscript{3} and racemic (top) vs. enantiopure (bottom) 1-phenylethylamine. The insets present the signal for the \textsuperscript{13}C spins denoted by the asterisks on the molecules. Both spectra were acquired at 298\,K by averaging 32 transients. The bottom spectrum shows a diastereomeric splitting, where the distinct peaks correspond to the starred carbon nuclei in both L-1-phenylethanol and D-1-phenylethanol. These starred nuclei are usually equivalent due to fast rotation of the phenyl group. Note the omission of the hydrogen atom indicates a racemic mixture of molecules.}
  \label{Fig2}
\end{figure}

\indent Physical processes mediated by the weak interaction are known to be asymmetric with respect to spatial inversion due to the parity nonconserving (PNC) nature of the weak interaction. Parity nonconservation was initially suggested by Lee and Yang in 1956 \cite{lee1956question}, and the first experimental demonstration was provided by Wu et al in 1957, by observing the $\beta$-decay of oriented \textsuperscript{60}Co nuclei \cite{wu1957experimental}. Further experimental proof was found in an asymmetry in the emitted electron spin polarization during muon decays \cite{garwin1957observations}. These initial demonstrations were at a nuclear level, but PNC effects are also observable at the atomic level, in the electronic spectra of atomic systems. The weak interaction between the electrons and the nucleons allows for nonvanishing electric-dipole transition amplitudes between states of nominally the same parity. This has been demonstrated by the observation of optical-rotation and Stark-interference experiments with heavy atoms: see \,\cite{bouchiat2011atomic} and references therein.\\
\indent Parity violation effects should also be observable in molecular systems, and many experiments have been proposed to detect molecular PNC. One method is to compare transition frequencies in the spectra of chiral enantiomers, and this has been attempted with microwave, infrared, and M{\"o}ssbauer spectroscopy\,\cite{bauder1997combined,kompanets1976narrow,lahamer2000search,darquie2010progress,crassous2003search}. Other recently explored methods are to observe time-dependent optical activity in chiral molecules\,\cite{szabo1999demonstration,macdermott2004proposed}, or using Stark interference techniques in diatomic molecules\,\cite{demille2008using,cahn2014zeeman}. These attempts have so far not measured molecular PNC.\\
\newline
\indent In this paper we explore an alternative method to observe molecular PNC, utilizing nuclear magnetic resonance (NMR) spectroscopy\,\cite{robert2001nmr,barra1986parity,barra1988possible,laubender2003ab,soncini2003parity,weijo2005perturbational,bast2006parity,nahrwold2014parity} to detect a chemical shift\footnote{Chemical shift, $\BD$, is defined as the magnetic shielding, $\BS$, relative to some reference: $\BD=\BS-\BS_{\rm{ref}}$.} difference between two enantiomers. In conventional NMR experiments, neglecting PNC effects, we expect to see identical chemical shifts for the nuclei of two enantiomers. However, if the PNC contribution to the nuclear spin Hamiltonian is considered, the chemical shifts of the nuclei in the two enantiomers are no longer equal. This parity-violating chemical shift difference arises because the electronic environments experienced by the nuclei are slightly asymmetric due to coupling with the chiral weak force. This is illustrated in Figure~\ref{Fig1}. In particular, PNC interactions will contribute to the magnetic shielding tensor $\BS$, the indirect spin-spin coupling tensor $\mathbf{J}$, and to the spin-rotation tensor $\mathbf{M}$\,\cite{barra1986parity}. As suggested by Bouchiat and Bouchiat\,\cite{bouchiat1997parity}, experimental searches for P-odd effects in the aforementioned parameters in molecules would result in the first observations of parity-violating forces in a static system, in contrast to atomic PNC experiments which involve transition processes.\\
\indent In this work we focus on NMR spectroscopy, and as such, we are concerned with the \textit{nuclear spin-dependent} (NSD) PNC contributions to the magnetic shielding tensor $\BS$. In particular, the magnetic shielding tensor for nuclei in a chiral molecule can be written as follows:
\begin{equation}
\BS_{\pm}=\BS^{_{\rm{\small{PC}}}} \pm \BS^{_{\subPNC}},
\end{equation}
where $\BS^{_{\rm{\small{PC}}}}$ denotes the parity conserving part of the magnetic shielding tensor, and $\BS^{_{\subPNC}}$ the parity nonconserving part which has precisely the same absolute values but opposite sign for the two enantiomers (right- and left-handed) of a chiral molecule. This results in equal but opposite (observed) NMR frequency shifts for the right- and left-hand isomers of a chiral molecule, or in an observed NMR spectral line splitting $\Delta\nu_{_{\rm{PNC}}}=(\gamma/\pi)\sigma^{_{\subPNC}}B_0$ in a racemic mixture of the chiral molecule in an isotropic solution. $B_0$ denotes the magnitude of the static magnetic field of the NMR spectrometer, $\gamma$ is the nuclear gyromagnetic ratio, and $\sigma^{_{\subPNC}}$ is the isotropic shielding factor, spherically averaged over all molecular orientations. In this work we primarily consider the isotropic shielding factor.\\
\indent NSD-PNC effects arise predominantly due to the following interactions\,\cite{ginges2004violations,khriplovich1991parity}: neutral weak-current interactions between the electron and the nucleus\,\cite{novikov1977translation}; electromagnetic interactions of the electron with the nuclear anapole moment\,\cite{flambaum1980translation}; and spin-independent electron-nucleon weak interactions combined with magnetic hyperfine interactions\,\cite{flambaum1985translation}. These effective NSD P-odd interactions of electrons with nucleons, relevant in an NMR experiment through their contribution to $\BS^{_{\subPNC}}$, can be written (in the non-relativistic limit) as follows:
\begin{align}
\begin{split}\mathcal{H}^{\rm{eN}}_{\rm{\small{PNC}}}&=\frac{G_{\rm{F}}\,\alpha^2}{2\sqrt{2}m}(\mathcal{H}^{(2)}_{\rm{\small{PNC}}}+\mathcal{H}^{(3)}_{\rm{\small{PNC}}}) \\
&=\frac{G_{\rm{F}}\,\alpha^2}{2\sqrt{2}m}\big\{ g_{2} \rm{\bf{I}}\left[\rm{\bf{p}},\delta(\rm{\bf{r}}-\rm{\bf{R}})\right]_{-} +i {g}_{3}\left[\mathbf{s}\times \mathbf{I} \right] \left[\rm{\bf{p}},\delta(\rm{\bf{r}}-\rm{\bf{R}})\right]_{+}\big\},
\end{split}
\label{eq:H}
\end{align}
where $G_{\rm{F}}$ is the Fermi constant ($G_{\rm{F}}=1.16637\times10^{-11}$\,MeV$^{-2}$), $\alpha$ is the fine structure constant, $m$, $\mathbf{p}$, and $\mathbf{s}$ are the mass, momentum and spin of the electron, respectively, $\mathbf{I}$ is the nuclear spin operator, and $\rm{\bf{r}}$ and $\rm{\bf{R}}$ are the coordinates of the interacting electron and nucleon, respectively (the delta function signifies the short range of the P-odd interaction).\\
\indent The coupling constants $g_2$ and $g_3$ represent the strengths of the P-odd interactions between the electrons and nucleus. To observe the $g_3$ contribution from ${\cal H}^\mathrm{eN}_\mathrm{PNC}$ to the shielding tensor, the spin-orbit interaction must be present (resulting in the mixing of singlet and triplet states)\,\cite{barra1986parity,gorshkov1982p}. The strength of the interactions proportional to the coupling constants $g_2$ and $g_3$ scale as $Z^2$ and $Z^2(\alpha Z)^2$ (where $Z$ is atomic number), respectively, because the PNC effect scales with electron spin density at the nucleus, and the electron momentum. For heavy nuclei ($Z\sim 80$) we need to use the relativistic operator instead of Eqn.\,\ref{eq:H}, which adds an additional $R(Z)$ factor, where $R(80)\sim 10$. Both terms of the non-relativistic operator (2) arise from one relativistic operator with a single coupling constant, so we may expect that $g_2 \approx g_3$.\\
\indent It is important to note that in the coupling constants $g_2$ and $g_3$, effects due to PNC-interactions of the molecular electrons with the nuclear anapole moment are also included, which in addition to the $Z^2$ dependence, will lead to a $g_2\,,g_3 \sim A^{2/3}$ dependence (where $A$ is the mass number of the nucleus under investigation). We note here that the $Z^5$ scaling law typical in PNC searches using chiral molecules arises due to nuclear spin-independent PNC interactions\,\cite{quack2008high}, to which we are not sensitive when measuring PNC contributions to the nuclear shielding tensor.\\
\indent It is clear from the scaling with $Z$ and $A$ that expected PNC effects can be increased by choosing molecules containing heavy nuclei. For heavy nuclei ($Z\sim 80$), by including the relativistic scaling, the expected energy shift is $\sim g_2 \sim 10^{-2}$ Hz. Theoretical calculations have been performed for different chiral molecules containing light and heavy nuclei, such as $^1$H, $^{13}$C, $^{195}$Pt, $^{203}$Tl, $^{205}$Tl and $^{207}$Pb, and the estimated principal values for the $\sigma^\mathrm{PNC}$ tensor can range from $10^{-10}$ ppm for light nuclei to as large as $10^{-5}$ ppm for heavy ones, corresponding to around $10^{-8}$ Hz to $10^{-3}$ Hz for the nuclei, considered at a static field strength of 20\,T\,\cite{barra1986parity,barra1988possible,laubender2003ab,soncini2003parity,weijo2005perturbational,bast2006parity,nahrwold2014parity}.\\
\newline
\indent The most straightforward way to see PNC effects in an NMR spectrum would be to observe a parity-violating chemical shift difference, manifesting as two distinct peaks in the signal acquired from a racemic (equal amount of left- and right-handed enantiomers) mixture of chiral molecules. Unfortunately, spin coherence lifetimes are limited, producing spectral peaks intrinsically broad with respect to the frequency shifts we are trying to observe\,\cite{robert2001nmr}.\\
\indent There are, however, techniques used by NMR spectroscopists to differentiate between enantiomers in a racemic mixture\,\cite{pirkle1967nonequivalence}. For example, a chiral solvent may be added to create a chiral environment, as shown in Fig.~\ref{Fig8}. The solvent molecules have a specific ‘handedness’, and typically interact non-covalently with the analyte to form diastereomers. This means the nuclei in the enantiomers experience a unique time-averaged electronic structure, which shows up as a chemical shift difference\,\cite{wenzel2003chiral}. In this work we refer to this as a diastereomeric splitting, and the characteristic spectra are shown in Fig.~\ref{Fig2}. Perhaps the most common application of this technique is to quantify the enantiomeric purity of a solution\,\cite{wenzel2003chiral}.\\
\\
\indent We propose a new experiment in which the diastereomeric splitting of given nuclear spins in a racemic mixture of sensor molecules is measured as the enantiomeric excess of the chiral solvent is titrated from left- to right-handed, or vice versa. With increasing enantiomeric excess the sensor shows a larger diastereomeric splitting. As the titration proceeds through the racemic point, the parity-conserving diastereomeric splitting vanishes, but the parity-violating chemical shift difference remains, and manifests as a residual splitting. Measuring the diastereomeric splitting as a function of the enantiomeric excess of the solvent, and interpolating the data, allows us to extract with higher precision the parity nonconserving component of the chemical shift as the effective racemic residual splitting.
\section{Experimental procedures and results}
For a preliminary demonstration we use 1-phenylethanol as the racemic sensor in the solvent 1-phenylethylamine. These molecules were chosen due to their ready availability, miscibility, and the fact that their similar geometries are expected to maximize intermolecular interactions and thus enhance diastereomeric splittings. We worked with natural abundance (approximately 1.1\%) of the \textsuperscript{13}C spin isomer.\\ 
\indent In particular, 2.5\,mL 1-phenylethanol, 2.5\,mL L-1-phenylethylamine, and 2.5\,mL deuterated chloroform were mixed using a 5\,mL graduated syringe to form a stock solution. A second stock solution was made in the same way, but with D-1-phenylethylamine replacing L-1-phenylethylamine. The deuterated chloroform acts as a reference compound for the chemical shift in ${}^{13}$C NMR experiments, and the deuterium gives a signal to which the magnetic field can be locked to prevent field drift. Twenty-one samples (each of volume 0.6\,mL) were prepared by mixing the stock solutions in varying ratios, in 30\,$\mu$L steps, using a 1\,mL graduated syringe. The solutions were mixed directly into 5\,mm NMR tubes, which were immediately sealed with parafilm. All NMR experiments were performed on a Bruker Avance III 850\,MHz spectrometer with a 5\,mm TXI liquid-state probe.\\
\begin{figure}
  \includegraphics[width=\columnwidth]{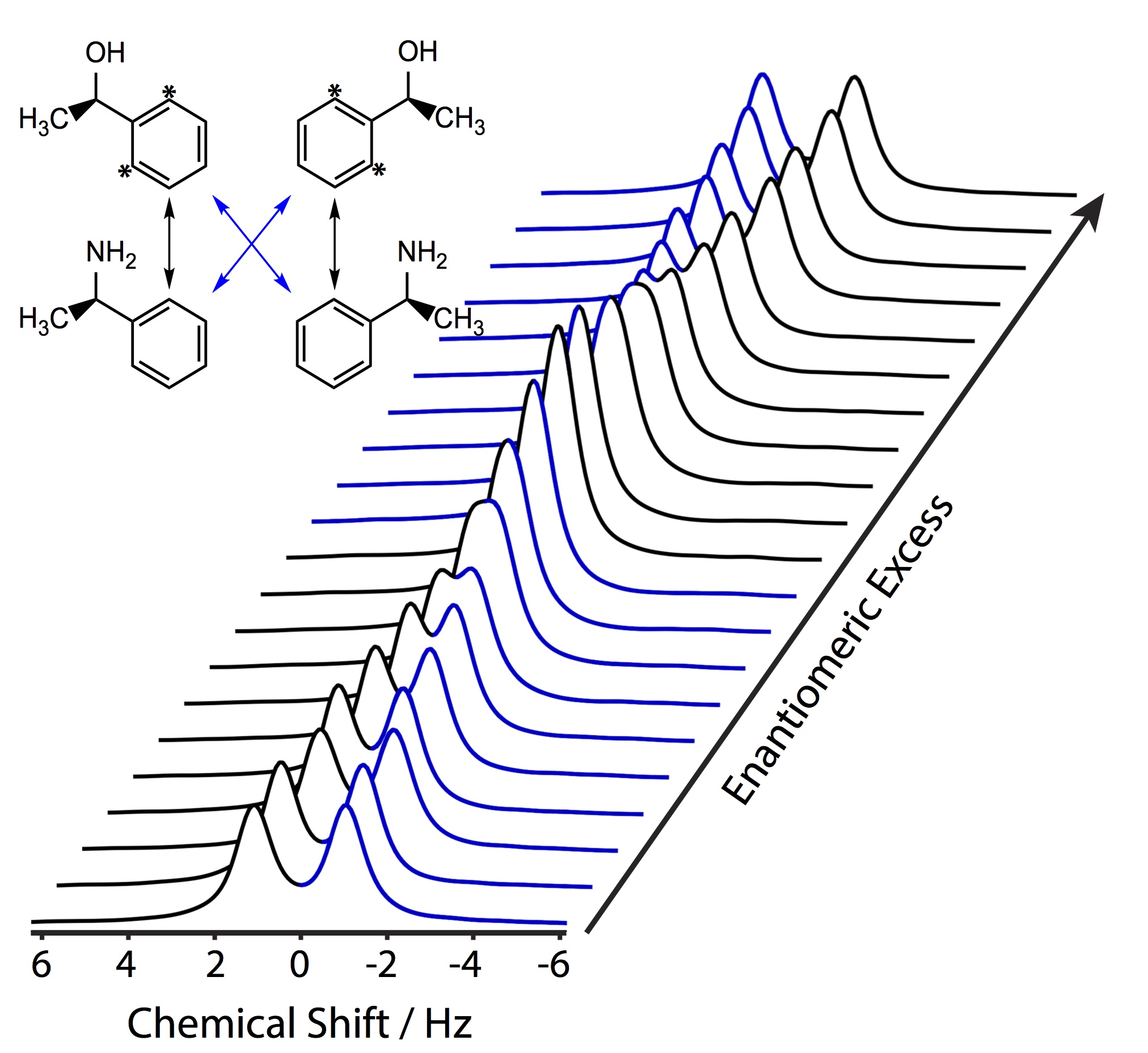}
  \caption{Stacked \textsuperscript{13}C spectra showing diastereomeric splitting of 1-phenylethanol as the enantiomeric excess of the 1-phenylethylamine environment is varied from 100\% D to 100\% L. The scale is in hertz, and centered on the peak of interest. All \textsuperscript{13}C spectra were acquired by averaging 32 transients, with proton decoupling, and have line broadening\,\cite{keeler2011understanding} of 0.5\,Hz applied. The inset shows the four possible diastereomeric interactions between the sensor (1-phenylethanol) and chiral solvent (1-phenylethylamine).}
  \label{Fig3}
\end{figure}
\indent In this experiment we compare the diasteriomeric splittings for \textsuperscript{1}H and \textsuperscript{13}C nuclei in order to look for an anomalous additional \textsuperscript{13}C splitting that would arise due to the $\sim$Z\textsuperscript{2} scaling of the PNC effects, which should result in a substantially larger effect for \textsuperscript{13}C than for \textsuperscript{1}H. Another way to consider this experiment is that the \textsuperscript{1}H spectra provide a measure the enantiomeric excess of the chiral solvent in each solution. We are therefore insensitive to uncertainty in the ratio of solvent enantiomeric excess in each solution, and are limited instead by the frequency precision of the NMR measurements.\\
\indent For each sample, the diastereomeric splitting of the proton directly bound to the chiral carbon in the sensor 1-phenylethanol was measured at 298\,K and 20\,T, using the average of 8 scans at 3.9\,s each. The \textsuperscript{1}H signals were fit to Lorentzian lineshapes, and the peak centers were specified with a precision of better than 40\,mHz. The diastereomeric splitting of the ortho carbon was measured at 298\,K and 20\,T, using the average of 32 scans at 3.1\,s each. The \textsuperscript{13}C signals were fit to Lorentzian lineshapes, and the peak centers were specified with a precision of better than 10\,mHz.\\
\indent In Fig.\,\ref{Fig3} we present the proton-decoupled \textsuperscript{13}C spectra as a function of the enantiomeric excess of the 1-phenyethylamine solvent. We observe a linear dependence of the induced diastereomeric splitting. Because the sensor molecules are in a racemic mixture, the relative peak integrals are independent of the enantiomeric excess of the chiral solvent. One peak signifies the presence of left-left and right-right diastereomers, and the other signifies left-right and right-left, as illustrated in the inset.\\
\indent In Fig.~\ref{Fig4} we show a plot of the \textsuperscript{13}C diastereomeric splitting as a function of the \textsuperscript{1}H diastereomeric splitting. We include a linear fit, derived through Deming regression\,\cite{deming1943statistical} accounting for errors in both variables, and this yields a residual splitting at the racemic point, where the \textsuperscript{1}H diastereomeric splitting is zero, of $1.8\,\pm\,3.4$\,mHz, or $(2.1\,\pm\,4)\times10^{-6}$\,ppm. We therefore do not observe a residual splitting within one standard deviation of our experimental precision.\\
\indent For the sensor molecule used in this study, we expect a \textsuperscript{13}C PNC chemical shift on the order of $10^{-10}$ to $10^{-8}$\,ppm, corresponding to approximately $\sim$\SI{20}{\nano\hertz} to \SI{2}{\micro\hertz} for a magnetic field of 20\,T. Clearly, our current experimental precision is insufficient to resolve this for a ${}^{13}$C nucleus. However, based on the calculations presented in Refs.\,\cite{barra1986parity,barra1988possible}, the value for high-Z nuclei like ${}^{129}$Xe \cite{bartik2001molecular,ruiz2006diastereomeric,jeong2015investigation}, $^{195}$Pt, $^{203}$Tl, $^{205}$Tl and $^{207}$Pb, is estimated to be 4-5 orders of magnitude larger. Therefore, an experiment using one of these nuclei and achieving the same measurement precision as demonstrated here may be able to observe molecular parity violation.\\
\begin{figure}
  \includegraphics[width=\columnwidth]{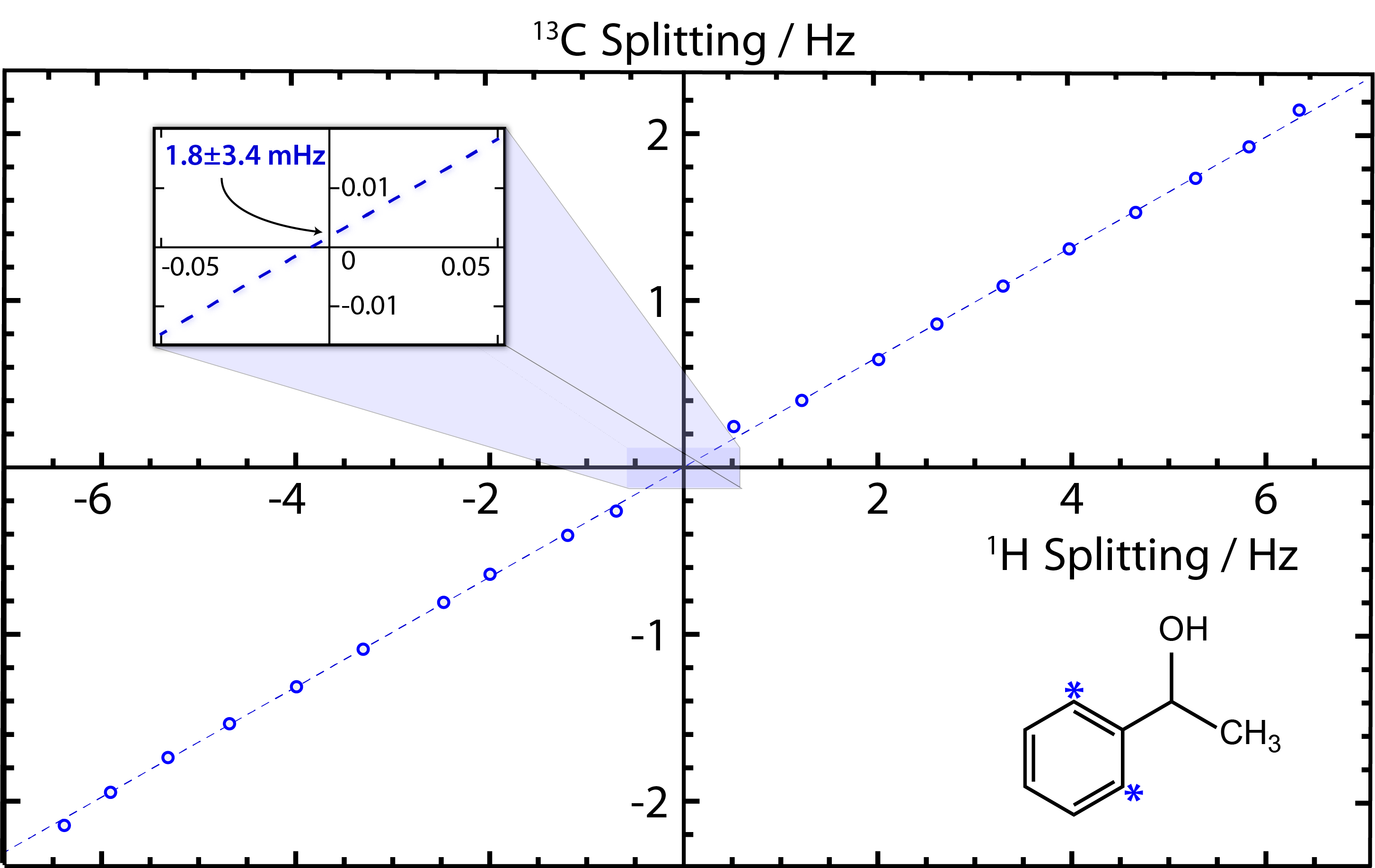}
  \caption{Experimental data showing the diastereomeric splitting of the \textsuperscript{13}C peaks vs. the diastereomeric splitting of the \textsuperscript{1}H peaks as the chirality of the solvent is titrated. A nonzero intercept on the vertical axis would be an indication of PNC effects.}
  \label{Fig4}
\end{figure}
\section{Discussions}
We now discuss various crucial aspects of the experiment and propose directions towards NMR detection of molecular parity violation. There are two parallel avenues towards sensitive measurement of the parity nonconserving chemical shift difference: increasing the magnitude of the effect by selecting appropriate molecules (for example, choosing a sensor molecule with a high-Z atom with nuclear spin 1/2), and increasing the precision of the experimental technique (for example, increasing signal-to-noise with higher spin polarization and higher static field strength).
\subsection{Increasing PNC Splitting - Z-Dependence}
As discussed previously, the $\BS^{_{\subPNC}}$ tensor (and its trace) is strongly dependent on atomic number, scaling faster than Z\textsuperscript{2}. For this reason, the ideal experiment would involve measuring the chemical shifts of chiral heavy-metal complexes. Platinum, thallium and lead are appealing, because they all have spin-1/2 nuclei. Here we do not consider nuclei with spin $>$1/2, because the nuclear quadrupole moment enables rapid relaxation, leading to broad spectral peaks.\\
\subsection{Orientation Dependent PNC - Solvent Anisotropy}
\indent The magnetic shielding is a tensor property, so \textsuperscript{13}C chemical shifts generally depend on molecular orientation with respect to the static external field ($B_0$), an effect known as chemical shift anisotropy (CSA). This is important because it is possible for the parity nonconserving part of the magnetic shielding tensor to be substantially anisotropic. For example, one can imagine a case where $\BS^{\rm{PNC}}$ is traceless, i.e., 
\begin{equation}
\sigma^{\rm{PNC}}_{\rm{iso}}=\frac{1}{3}\left(\sigma^{\rm{PNC}}_{xx}+\sigma^{\rm{PNC}}_{yy}+\sigma^{\rm{PNC}}_{zz}\right)=0,
\label{test}
\end{equation}
making it impossible to measure PNC using our proposed method in isotropic solutions.
By aligning the molecules and thus enhancing the contribution of one (or more) of the $\BS^{\rm{PNC}}$ components, the PNC contribution to the observable spectrum can be maximized.
Common alignment techniques for NMR applications include dissolving molecules in liquid crystalline solvents \cite{Emsley1975} or stretched polymer gels \cite{Tycko2000}.
The residual magnetic shielding tensor ($\BS_{res}$), which is the observable in an experiment, may be related to the shielding by way of an alignment tensor, $\boldsymbol{A}$ where $A_{ij}$ are the matrix elements of the molecular alignment tensor. Typical values of $|A_{ij}|$ for molecules dissolved in a liquid crystal at room temperature can be $10^{-4}-10^{-1}$.\\
\indent An important consideration when using a liquid crystal solvent is the increased viscosity of the solution, which has the effect of slowing molecular tumbling. This affects the $T_2$ relaxation time, which can increase linewidths, and one loses spectral resolution. In our demonstration, by using a PBLG solvent the peak splitting increased by a factor of $\sim$6, but the peak linewidths increased by a factor of $\sim$8. Overall the measurement precision has therefore decreased, and further optimization is required to justify the use of a liquid crystal solvent.
\subsection{Increasing NMR Precision - Field Strength}
\indent In these experiments we measure chemical shift differences, which depend linearly on static field strength, $B_0$. We can therefore increase the magnitude of the splitting, and hence enhance the precision of our measurements, by using a higher static field.\\
\indent Another benefit of using a high $B_0$ field is an improvement to the signal-to-noise ratio (S/N) of the peaks. For a conventional high-field experiment on a nonconducting sample, S/N depends on field strength as $(B_0)^\frac{7}{4}$\,\cite{hoult2007sensitivity}. The uncertainty with which we can specify the center of a peak with a certain width is proportional to the inverse of the signal-to-noise ratio.\\
\indent We must also consider the effect of static field drift over time. Field drift during signal acquisition can broaden the spectral lines and reduce the measurement precision, but this effect is almost completely negated by locking to a deuterium nucleus. A typical value for field drift over the course of a day on a modern high-field spectrometer might be a few parts per billion, or a few hertz. Because we plot \textsuperscript{1}H chemical shifts against \textsuperscript{13}C chemical shifts, field drift between acquiring the \textsuperscript{1}H and \textsuperscript{13}C spectra is a systematic error that might be mistaken for observation of molecular PNC. In our case, the \textsuperscript{13}C spectra show a chemical shift difference between the carbons of up to $\sim$2\,Hz, or $\sim$2\,ppb. We can therefore expect a shift between the peaks, over the course of a day, of perhaps 2 parts per quintillion, or 2\,nHz.\\
\indent The real time between the \textsuperscript{1}H and \textsuperscript{13}C experiments on a sample can be reduced to a few seconds, and the drift can be reduced further by \textsuperscript{2}H-locking. If field drift still appears as a significant error in the final plot, it is possible to use a spectrometer that allows for simultaneous multi-channel acquisition.
\subsection{Increasing NMR Precision - Hyperpolarization}
\indent The precision of a measurement improves with increasing signal-to-noise ratio. This was demonstrated in the field of enantiodiscrimination by Monteagudo et al.\,\cite{monteagudo2017chiral}, who used dissolution dynamic nuclear polarization\,\cite{ardenkjaer2003increase} (d-DNP) as a way of greatly increasing the S/N of their experiment. Typical signal enhancements of \textsuperscript{13}C nuclei in small organic molecules are in the order of $10^3-10^4$. It should be possible to hyperpolarize the sensor molecule using DNP at low temperature, rapidly dissolve the sample using the chiral solvent at a specific enantiomeric excess, flow this mixture into the NMR magnet, and acquire the \textsuperscript{1}H and \textsuperscript{13}C spectra before the spin polarization returns to thermal equilibrium.\\
\indent Currently the precision of our measurement is approximately three to four orders of magnitude less than required to observe PNC line shifts at the expected magnitude of $\sim$\SI{20}{\nano\hertz} to \SI{2}{\micro\hertz} for \textsuperscript{13}C. Polarizing \textsuperscript{1}H and \textsuperscript{13}C nuclei via d-DNP could enhance the signal-to-noise in both axes by a factor of up to 10\textsuperscript{4}, and hence the overall frequency precision by 10\textsuperscript{4}. A further 10\textsuperscript{2} signal-to-noise enhancement can be achieved in the \textsuperscript{13}C spectra by isotopically enriching the sensor molecule with \textsuperscript{13}C nuclei (compared to the natural abundance of 1.1\% \textsuperscript{13}C present in the unenriched compounds used in our experiments). This further enhances the measurement precision in the \textsuperscript{13}C axis, and hence the overall precision. In this work we acquired the \textsuperscript{13}C spectra using 32 transients, and the \textsuperscript{1}H spectra using 8 transients, whereas dissolution-DNP is a single-shot experiment. One has to factor in the reduced duty cycle. Overall, by employing d-DNP and observing similar molecules, a precision enhancement by a factor of $\sim$10\textsuperscript{3}-10\textsuperscript{4} might be achieved. Other contributing factors might decrease this figure, such as the radicals needed for hyperpolarization inducing relaxation of the nuclei, leading to broader spectral lines. However, overcoming the infamous low sensitivity in the NMR experiment must be a serious consideration in the search for PNC effects.\\
\indent Another potential hyperpolarization modality is parahydrogen-induced polarization (PHIP) \cite{bowers1987parahydrogen}, which can provide similar enhancements to DNP. Non-hydrogenative PHIP (also called signal amplification by reversible exchange, or SABRE) \cite{adams2009reversible} has the additional advantage of providing a continuously hyperpolarized sample \cite{hovener2013hyperpolarized}, and has been demonstrated for relatively heavy nuclei like ${}^{119}$Sn \cite{olaru2016using}. It would also be interesting to explore whether coherent experiments, like the NMR ``RASER'' of Ref.~\cite{suefke2017hydrogen}, could be beneficial for measuring PNC.\\
\subsection{Condensed Matter Experimental Challenges}
Condensed matter experiments of this style might pose challenges. Even if a residual diasteriomeric splitting at the racemic point is detected, it is potentially unclear whether this is a signature of parity violation, or if it results from a chiral impurity in the solution.\\
\indent If the chiral impurity in the experiments is a miscalibrated concentration of the chiral solvent (in this case, 1-phenylethylamine), this should appear in both the \textsuperscript{1}H and \textsuperscript{13}C diastereomeric splitting measurements and not affect the fitting. However, if the impurity is any other chiral molecule, such as an amino acid or other biological contaminant, for example, from human contact, this might produce a different shift for the \textsuperscript{1}H and \textsuperscript{13}C sites, and thus mimic PNC effects. For this experiment to be a viable method of measuring PNC effects, the solutions should be prepared in a contaminant-free environment, and the NMR tubes should be sealed to prevent contamination.\\
\indent To estimate the magnitude of these effects, we note that a diastereomeric splitting of $\sim$2\,Hz is observed on our sensor molecule 1-phenylethanol in a solution of 2.59\,M D- or L-1-phenylethylamine, the chiral solvent. If we assume the worst-case where the chiral contaminant induces an identical diastereomeric splitting on the \textsuperscript{13}C nucleus, and none on the \textsuperscript{1}H nucleus, it would have to be in $\sim$1\,nM concentration to induce a 1\,nHz splitting on the \textsuperscript{13}C site. This concentration should be detectable with high sensitivity methods such as mass spectrometry.
\section{Conclusion}
We have proposed a new procedure to search for parity nonconservation in the chemical shifts of chiral molecules using NMR spectroscopy. We demonstrate this experiment as a proof-of-principle, and set an upper limit on the parity nonconserving chemical shift difference between the ortho carbon nuclei in 1-phenylethanol enantiomers of $1.8\,\pm\,3.4$\,mHz, at a static field strength of 20\,T. The expected size of this effect for the studied system is on the order of $10^{-8}-10^{-6}$\,Hz.
For an ideal experiment in which the target molecule is a chiral $^{207}$Pb complex, we estimate that the parity violating chemical shift difference will be on the order of $10^{-3}$\,Hz, which may make the PNC contribution accessible to NMR experiments with existing high-field NMR equipment.
Even if experiments are unable to reach limits corresponding to PNC predicted by the Standard Model, precision measurements of this kind may be useful for testing exotic physics models that predict the presence of parity-violating cosmic fields\,\cite{roberts2014parity,roberts2014limiting}.
\section{Acknowledgments}
We thank Drs. Robert Graf, Manfred Wagner and Manuel Braun at the Max Planck Institute for Polymer Research for providing instrument access for these experiments. We thank Prof. Oleg P. Sushkov for helpful discussions. This project has received funding from the European Research Council (ERC) under the European Union’s Horizon 2020 research and innovation programme (grant agreement No 695405). We acknowledge the support of the Simons and Heising-Simons Foundations, and the DFG Reinhart Koselleck project. James Eills acknowledges the DAAD for research funding. Lykourgos Bougas acknowledges support by a Marie Curie Individual Fellowship within the second Horizon 2020 Work Programme. Mikhail Kozlov acknowledges the Helmholtz-Institut, Mainz, and the MITP visitor programs. We would also like to thank Dr. Clancy Slack and Ashley Truxal for enlightening discussions about enantiomeric  discrimination by hyperpolarized ${}^{129}$Xe chemical shifts.

\bibliographystyle{unsrt}
\bibliography{PNC_References}{}

\end{document}